\documentclass[conference]{IEEEtran}

\newif\ifcameraready

\camerareadytrue

\usepackage{cite}
\usepackage{amsmath,amssymb,amsfonts}
\usepackage{algorithmic}
\usepackage{graphicx}
\usepackage{textcomp}
\usepackage{xcolor}
\usepackage{subcaption}
\usepackage{xspace}
\usepackage{booktabs}
\usepackage{hyperref}
\usepackage{multirow}
\usepackage{tipa}
\def\BibTeX{{\rm B\kern-.05em{\sc i\kern-.025em b}\kern-.08em
    T\kern-.1667em\lower.7ex\hbox{E}\kern-.125emX}}
\ifcameraready
    \newcommand{\A}{{Ahoy}\xspace}
    \newcommand{\B}{{Pulse}\xspace}
    \newcommand{\C}{{IDE}\xspace}
    \newcommand{\D}{{Arch}\xspace}
    \newcommand{\Dlong}{{Architecture ({Arch})}\xspace}
\else
    \newcommand{\A}{\texttt{A}\xspace}
    \newcommand{\B}{\texttt{B}\xspace}
    \newcommand{\C}{\texttt{C}\xspace}
    \newcommand{\D}{\texttt{D}\xspace}
    \newcommand{\Dlong}{\texttt{D}\xspace}
\fi

\begin{document}

\title{A Semi-Spontaneous Dutch Speech Dataset for Speech Enhancement and Speech Recognition}

\ifcameraready
    \author{
        \IEEEauthorblockN{
            Dimme de Groot\textsuperscript{*},
            Yuanyuan Zhang\textsuperscript{*},
            Jorge Martinez,
            Odette Scharenborg 
        }
        \IEEEauthorblockA{\textit{Multimedia Computing Group} \\
        \textit{Delft University of Technology}\\
        Delft, the Netherlands \\
        \{d.c.c.j.degroot, y.zhang-44, j.a.martinezcastaneda, o.e.scharenborg\}@tudelft.nl
    }
    
}

\fi

\maketitle
\begingroup
\renewcommand{\thefootnote}{*}
\footnotetext[1]{These authors contributed equally.}
\endgroup

\begin{abstract}
We present DRES: a 1.5-hour Dutch realistic elicited (semi-spontaneous) speech dataset from 80 speakers recorded in noisy, public indoor environments. DRES was designed as a test set for the evaluation of state-of-the-art (SotA) automatic speech recognition (ASR) and speech enhancement (SE) models in a real-world scenario: a person speaking in a public indoor space with background talkers and noise. The speech was recorded with a four-channel linear microphone array. 
In this work we evaluate the speech quality of six well-known single-channel SE algorithms and the recognition performance of eight SotA off-the-shelf ASR models before and after applying SE on the speech of DRES. We found that five out of the eight ASR models have word error rates (WERs) lower than 22\% on DRES, despite the challenging conditions. In contrast to recent work, we did not find a positive effect of modern single-channel SE on ASR performance, emphasizing the importance of evaluating in realistic conditions.
\end{abstract}

\section{Introduction}
Automatic speech recognition (ASR) systems are widely used in diverse scenarios, including emergency centres~\cite{valizada2021development}, health care services~\cite{johnson2014systematic}, and education~\cite{van2016evaluating}. To ensure reliable performance in practice, it is crucial that ASR systems are able to recognize speech from diverse speakers and under diverse acoustic conditions, including noisy environments with background talkers. However, most speech datasets for developing or evaluating noise-robust ASR and speech enhancement (SE) algorithms consist of synthetic noisy speech, created by mixing clean recordings with real or artificially generated noise~\cite{valentinibotinhao16_ssw, hu2007subjective, richter24_interspeech, hirsch00_asr, barker2013pascal, segura2007hiwire}. 
Despite their obvious usefulness, such synthetic noisy speech does not fully capture the complex acoustic characteristics of real noisy speech.

Real noisy speech better reflects real-world acoustic conditions, not only because it captures the intricate and time-variant noisy and reverberant conditions, but also partly because speakers naturally modify their speech to maintain intelligibility, a phenomenon known as the Lombard effect~\cite{Brumm2011}. These modifications depend on, among others, communicative intent and the acoustic environment~\cite{Junqua1999, Lu2008, Garnier2010, UmaMaheswari2020}. To address this, the CHiME-3 dataset was created, containing English speech recorded in caf\'e, street junctions, public transport, and pedestrian areas~\cite{barker2015third}. Similarly, the MISP-Meeting dataset consists of real-world far-field Mandarin speech recorded in meeting and conference rooms, with pervasive background noise~\cite{hangchen-etal-2025-misp}. 
These datasets enable more accurate evaluation of ASR and SE systems in real-world noisy environments~\cite{fujimoto2019, Iwamoto2022}.  \looseness=-1

Nevertheless, real noisy speech datasets remain scarce~\cite{Dua2023}, including for Dutch. Most publicly available Dutch speech datasets, such as the Corpus Gesproken Nederlands (CGN)~\cite{oostdijk-2000-spoken}, Jasmin-CGN~\cite{cucchiarini2006jasmin}, and the Dutch portion of Common Voice 12~\cite{ardila2019common}, were recorded in quiet environments, occasionally with a few noisy samples. Consequently, the performance of state-of-the-art (SotA) ASR systems on Dutch is typically only evaluated on clean speech. For example, OpenAI's Whisper models~\cite{radford2023robust} were evaluated on Dutch speech from Jasmin-CGN and Common Voice 12~\cite{fuckner2023uncovering, raes2024everyone, zhang2026speech}. Meta's Wav2vec 2.0~\cite{baevski2020wav2vec} model was evaluated on Jasmin-CGN~\cite{fuckner2023uncovering}, while two Dutch ASR models trained from scratch were evaluated on Jasmin-CGN~\cite{feng2024towards}.

This paper answers the open question of how well SotA ASR systems perform on real noisy Dutch speech. To that end we designed and collected DRES: a Dutch Realistic Elicited (or \textit{semi-spontaneous}) Speech dataset, containing speech from 80  speakers speaking Dutch in a real-world scenario, i.e. public indoor spaces with background talkers and noise. We choose to collect elicited speech as this is more naturalistic than read speech. DRES is recorded using a four-channel linear microphone array, facilitating future research in multi-channel SE. To keep the scope of the present study focused, we consider only single-channel conditions.

Using the speech from DRES, we evaluated eight SotA ASR models: Google Chirp 3 and Telephony~\cite{zhang2023google}, the Microsoft Azure ASR model~\cite{microsoft_speech_models}, the Massive Multilingual Speech model~\cite{pratap2024scaling}, two Whisper models~\cite{radford2023robust} (Whisper-large-V3 and Whisper-large-V3-turbo), NeMo-nl~\cite{nvidia_fastconformer_nl, nvidiacanary2024}, and a CGN-pre-trained Conformer model~\cite{zhang_dutchcgnconformerfbank, zhang2026speech}.

Speech signals might be pre-processed using SE algorithms before inputting them into an ASR, with the aim of improving ASR performance~\cite{HaebUmbach2021, Huang2025a}. A known limitation of SE is that it can introduce artefacts which have been shown to harm the performance of traditional HMM-GMM~\cite{6296526} and hybrid DNN-HMM ASR models~\cite{fujimoto2019, Iwamoto2022, Wang2020, Ho2023} trained on clean speech, particularly in single-channel settings. However, in the past decade, ASR models have dramatically improved by training large models on a large amount of speech~\cite{radford2023robust, zhang2023google, microsoft_speech_models}. At the same time, SE algorithms have also improved, and recent work on artificially created noise-speech mixtures showed that modern single-channel SE, such as SGMSE+~\cite{Richter2023}, has improved end-to-end (E2E) ASR performance on noisy speech~\cite{Lemercier2023, Yang2026, Chao2024}. However, these findings were reported for English, and the extent to which they generalise across languages remains unclear \cite{Giraldo2025}. Therefore the second open question addressed in this paper is how single-channel SE influences the performance of SotA ASR systems for Dutch real-world noisy speech. To answer this, first we applied six single-channel SE algorithms, including two simple but still widely used algorithms (spectral subtraction and spectral noise gating)~\cite{Berouti1979, Sainburg2020} and four modern deep-learning-based algorithms (MetricGAN-OKD, two variants of SGMSE+, and SEMamba )~\cite{Richter2023, Chao2024, Shin2023}, to the speech from DRES, and then evaluated the performance of the eight ASR models on the enhanced Dutch speech. Moreover, we report the objective speech quality before and after SE. DRES is available under end-user license agreement (EULA) from \cite{DresDataset}. 

\section{The DRES corpus}

\subsection{Corpus design}
To elicit spontaneous speech and to ensure both lexical and phonetic diversity, three tasks were designed:
\begin{enumerate}
    \item \textit{Free speech:} speakers were instructed to talk freely on a topic  of their own choice or chosen from a preselected list. \looseness=-1
    \item \textit{Picture card:} speakers were instructed to randomly select a picture from 26 pre-prepared picture cards and describe their card or tell a short story fitting that card.
    \item \textit{Prompt card:} speakers were instructed to randomly select a topic from 26 pre-prepared prompt cards and talk about it. 
\end{enumerate}

\begin{figure}[ht]
\centering
\begin{subfigure}{0.28\columnwidth}
    \includegraphics[width=\textwidth]{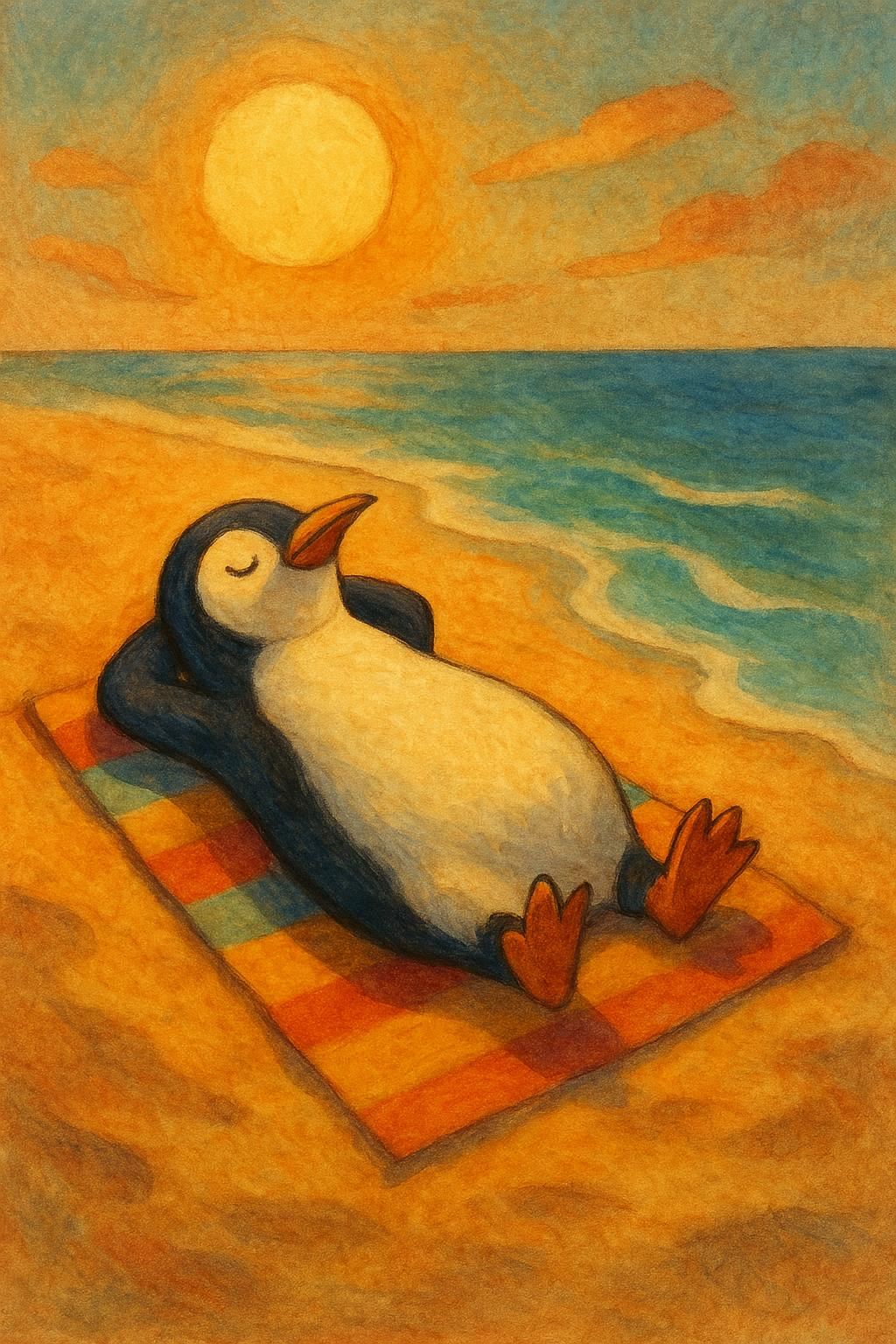}
    \label{pic_card}
\end{subfigure}
\begin{subfigure}{0.28\columnwidth}
    \includegraphics[width=\textwidth, trim={2.4cm 0cm 2.4cm 0cm}, clip]{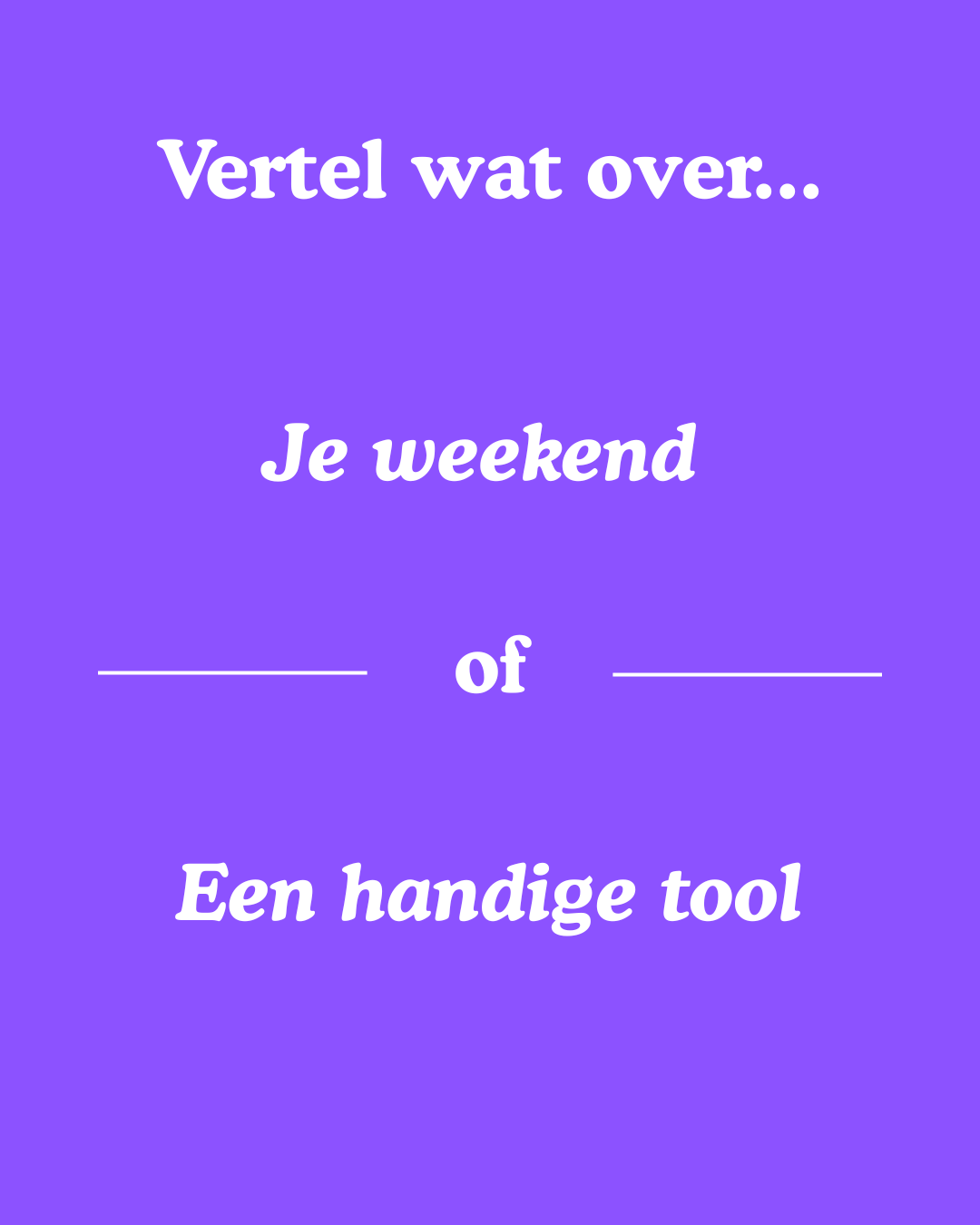}
    \label{prompt_card}
\end{subfigure}
\vspace{-10pt}
\caption{An example picture-card (left) and prompt-card (right). }
\label{fig_cards_and_mic}
\end{figure}

\noindent The prompt and picture cards were created by the 
\ifcameraready
first author (D.G.). 
\fi
Picture cards were designed to have a 
dreamlike aesthetic and were made with GPT-4o using the Bing image generation interface~\cite{openai_gpt4o_2024}. An example of a prompt-card and picture-card is shown in Figure~\ref{fig_cards_and_mic}. Prompt cards contained an instruction of the following form (translated from Dutch): ``Tell something about [topic 1] or [topic 2]". Example topics are ``your favourite season" or ``a dish you like to make".  

We first conducted a pilot study at \A (see Section~\ref{sec_2_2_rec}) with the \textit{free speech} elicitation task. 
Speakers were instructed to speak for at most 20 seconds into the microphones. From this pilot, the following limitations were identified: (1) 20 seconds was often insufficient, (2) speakers occasionally found it difficult to select a topic and (3) speakers frequently directed their gaze to the researcher instead of to the microphones.

Consequently, for the subsequent recording sessions (see Section \ref{sec_2_2_rec}), the recording design was modified: speakers were recommended to speak for maximally 45 seconds and were instructed to randomly select a prompt card and a picture card, with the option to redraw a card if desired. During recording, the experimenter 
\ifcameraready
(D.G.)
\fi
positioned himself behind the microphone array if space permitted. This led to a more natural setting for the speaker, and a better recording setting.  

\subsection{The speech recordings}\label{sec_2_2_rec}
The speech was recorded at the main halls of four different public buildings (\A, \B, \C, \Dlong
\ifcameraready
)
\fi 
to (1) recruit participants from a wide and diverse population, thus increasing speaker diversity and reducing the risk of participant re-identification, and (2) capture a variety of realistic indoor acoustic conditions. \A is an exhibition venue, comprising demonstration and exhibition areas in which multiple conversations were held simultaneously. \B, \C, and \D are university buildings. \B is a lunch area characterized by 
babble noise, \C is an open study area without enforced silence, and \D is an open creative space with ongoing activities. The recordings for \B, \C and \D were scheduled around lunchtime, which allowed for a sufficient number of participants, realistic noisy background conditions, and minimized disturbance to other people in the buildings. 

\begin{figure}[ht]
    \centering
    \includegraphics[width=0.7\columnwidth, trim={7cm 0cm 18cm 8cm}, clip]{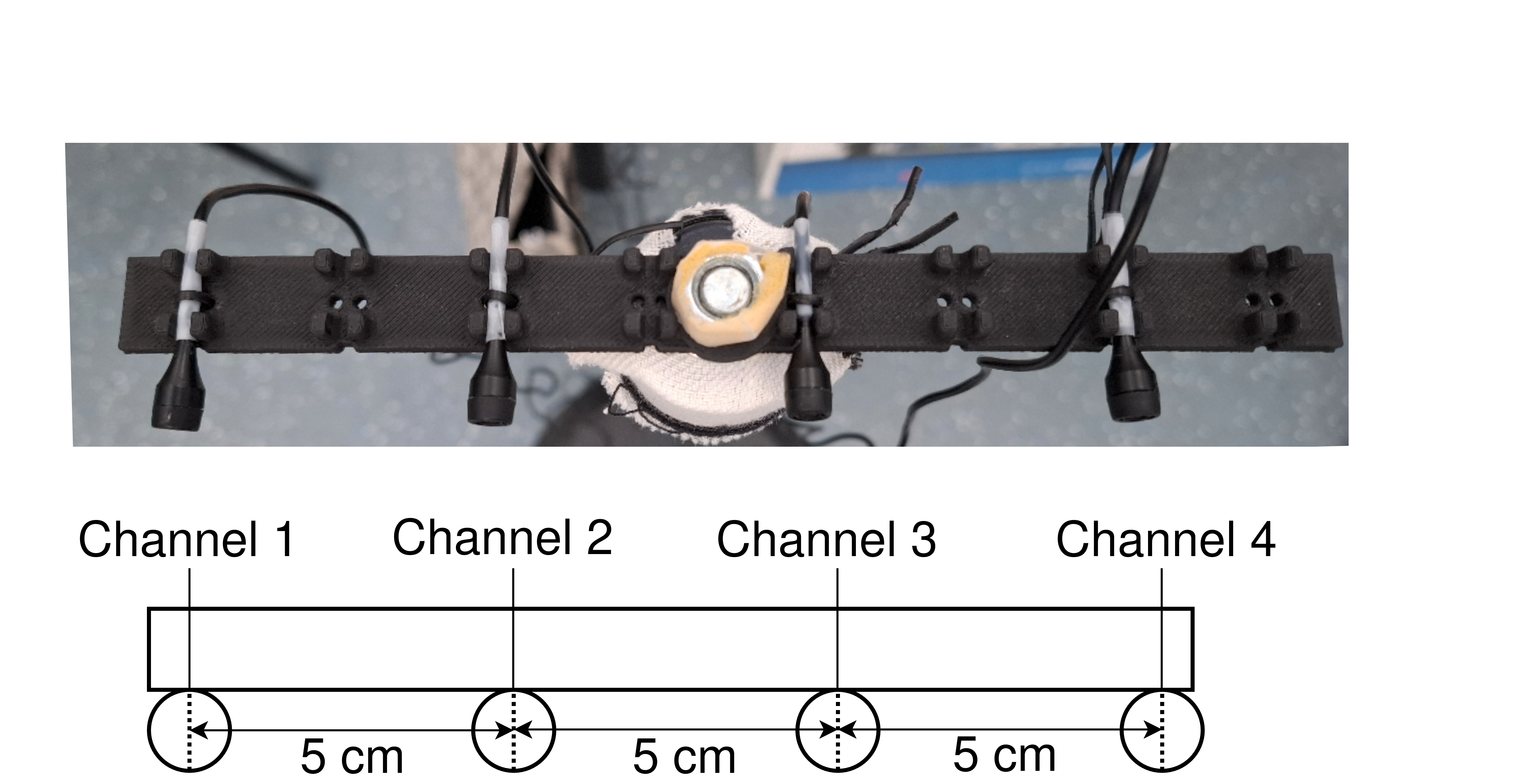}
    \caption{The four-channel microphone array used during the recording of DRES.}
    \label{fig_mic_array}
\end{figure}

All recordings were made using a linear microphone array consisting of four AKG C147 PP lavalier microphones with an inter-microphone spacing of 5 cm connected to an RME Fireface UFX III audio-interface recording at a sampling rate of 48 kHz. The microphone array is shown in Fig.~\ref{fig_mic_array}.
The exact distance of the speaker to the microphone-array was not controlled nor registered and was subjectively estimated to be 1.0 - 1.5 m in most cases. Data collection was approved by the Human Research Ethics Committee (HREC) from 
\ifcameraready
    Delft University of Technology,
\fi
including the informed consent form, the questionnaire for metadata collection, and the data management plan.

\subsection{Participants}
Participants were recruited on-the-spot by reaching out to people near the recording locations. Participants received no monetary compensation, and were only offered a small snack. In total, 80 speakers (65 native Dutch speakers, 12 non-native speakers, and 3 speakers that did not disclose) were recorded. Table~\ref{tab: A} lists the gender, age range, and nativeness of the speakers. 

\begin{table}[ht]
    \centering
    \caption{Gender, age and nativeness information of speakers at the recording locations.  Gender labels include: female (F), male (M), non-binary (N), other (O; genders other than female, male, and non-binary), and unknown (U; no disclosure). The rightmost column reports the total number of speakers and their nativeness (native$\vert$non-native$\vert$unknown).}\label{tab: A}
    \label{tab:speaker_stats}
    \resizebox{\columnwidth}{!}{
    \begin{tabular}{lcccccc|c} 
        \toprule
        \textbf{} & \textbf{18-30} & \textbf{31-40} & \textbf{41-50} & \textbf{51-60}  & \textbf{61-70} &\textbf{N/A} & \textbf{Total}\\ 
        \midrule
        \A & 2F, 3M, 1N  & 2F, 1M  & 2F  & 2F, 1M, 1O & 3F, 1M & 1U & 20 (17$\vert$2$\vert$1)\\
        \B & 8F, 13M  & - & -  & - & - & - &21 (16$\vert$5$\vert$0)\\
        \C & 10F, 10M  & -  & -  & - & - & - & 20 (18$\vert$1$\vert$1)\\
        \D & 6F, 11M, 2U & -  & -  & - & - & - & 19 (14$\vert$4$\vert$1)\\
        \bottomrule
    \end{tabular}
    }
\end{table}

\subsection{Transcriptions}
We created orthographic transcriptions following the protocol of Jasmin-CGN~\cite{cucchiarini2006jasmin}. 
For this, the raw audio recordings, with a duration ranging from 5.7 s to 93.3 s, were manually time-stamped into small chunks (0.4–24.8 s) by 
\ifcameraready
Y.Z.
\fi
\ifcameraready
Author D.G.
\fi
created the orthographic transcriptions. Three volunteers (all native Dutch speakers) checked and corrected the transcriptions when needed. DRES has a vocabulary size of 2,842 words, excluding non-linguistic symbols.
DRES consists of 1.5 hours of raw audio. After removing silences, 1.4 hours of speech was retained. The raw audio (prior to segmentation) was used for the SE experiments, while the segmented version was used for the ASR experiments.

\subsection{The quality of the speech data}
\label{subsubsec:SQ}
\begin{figure}[ht]
    \centering
    \ifcameraready
        {\includegraphics[width=0.85\columnwidth, trim={0cm 0.66cm 0cm 0.2cm}, clip]{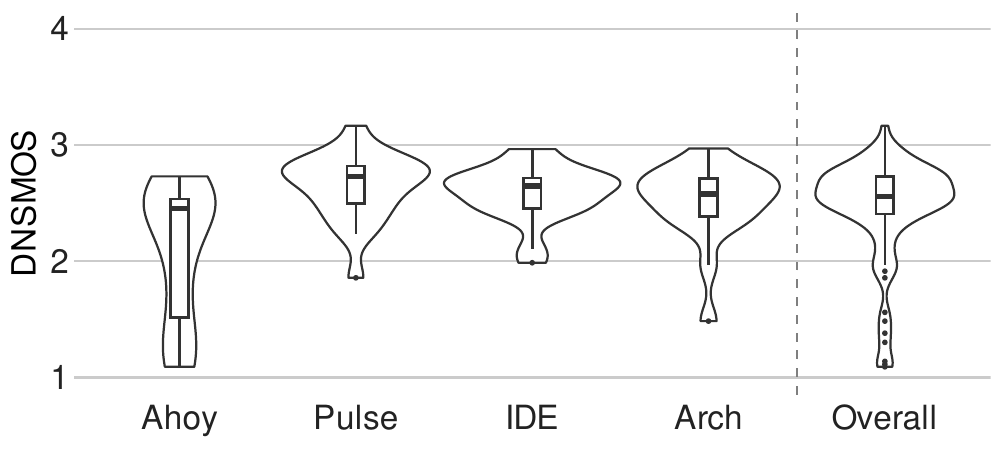}}
    \fi
    \caption{The distribution of the mean-opinion-score (MOS) for the recordings at the four recording locations, estimated using DNSMOS. Higher values imply a better quality. `Overall' represents the combined results over all locations.}\label{fig_quality_loc}
\end{figure}

As is mentioned in the introduction, although DRES is a multichannel dataset, in this work we focus on a single-channel analysis. For this we used the recordings from channel 2 of the microphone-array (see Fig. \ref{fig_mic_array}). Figure \ref{fig_quality_loc} shows the  distribution of the MOS scores, estimated using DNSMOS~P.835 \cite{Reddy2022} (see Section \ref{sec_eval}), of the speech signals per recording location and averaged over the four locations. The speech quality is particularly low for the recordings made at location \A, highlighting the difficult acoustic conditions. The quality is similar across the other three buildings. A Kruskal-Wallis test indicated that the DNSMOS differed over recording locations ($\mathcal{X}^2(3)=16.63, p<.001$). Post hoc pairwise comparisons were conducted using Dunn’s test with Holm-Bonferroni correction. DNSMOS scores were significantly lower for location \A than \B $\left(p<.01\right)$, \C $\left(p<.05\right)$ and \D $\left(p<.05\right)$. Comparisons between other locations were not significant.

\section{Experiments}
\subsection{Speech enhancement algorithms}\label{sec_se}
We selected six single-channel SE algorithms based on ease-of-use, computational complexity, SE performance, and popularity. The baseline speech signal and the SE algorithms are described below. 

As \textit{baseline signal} (Base) we use the recordings from channel 2, which is at the center of the microphone-array (see Fig. \ref{fig_mic_array}). Each utterance is resampled to 16 kHz and amplitude normalised to 0.707 to avoid clipping. All SE algorithms described below operate on these normalised and resampled utterances. 

\textit{Spectral subtraction} (SS) is a traditional and well-established low-complexity signal-processing-based SE technique which uses the difference between an estimate of the noise spectrum and the noisy-speech spectrum to compute the enhanced speech \cite{Berouti1979, Boll1979, OShaughnessy2024}. SS and variants thereof work well for stationary noise and have been used as baseline SE method for ASR in previous works (e.g.~\cite{fujimoto2019, Maganti2012}). We therefore include it as baseline.
We use the implementation in Pyroomacoustics \cite{Scheibler2018}.

\textit{Spectral Noise Gating} (SNG) is a low-complexity and general-purpose noise reduction method~\cite{Sainburg2020, Sainburg2021, Sainburg2025}. This algorithm class has previously been used as baseline for evaluating the effect of SE for ASR~\cite{Groot2025a} and for evaluating SE performance across languages~\cite{Giraldo2025}, where it outperformed SGMSE+ in terms of phoneme error rate and word information lost. We use the implementation \texttt{Noisereduce}~\cite{Sainburg2024}.

\textit{MetricGAN-OKD} (GAN) is a modern, causal, low-complexity SE method which uses a generative adversarial network and online-knowledge distillation to optimise the enhancement with respect to multiple perceptual metrics~\cite{Shin2023}. We use the implementation \texttt{enhanceSpeech} from Matlab R2025a.

\textit{SGMSE+} is a diffusion-based generative SE algorithm \cite{Richter2023}. SGMSE+ is a commonly used baseline and outperforms SotA discriminative SE algorithms in unseen conditions~\cite{Richter2023}. A disadvantage of SGMSE+ is that it is computationally expensive. We use both the pre-trained checkpoint on {WSJ0-CHiME3} (SG$_W$) and on {Voicebank-Demand} (SG$_V$) from the  publicly available implementation of SGMSE+ \cite{Welker2023git}. 

\textit{SEMamba} (SEM) is a popular predictive SE algorithm~\cite{Chao2024} which has previously been used for robust ASR~\cite{Wang2024}. We use the pre-trained checkpoint on Voicebank-Demand from the publicly available implementation of SEMamba~\cite{Chao2024git}.

\subsection{Automatic speech recognition models}
To capture the breadth of SotA ASR systems, we selected seven popular models which perform well on Dutch speech developed by Google, Microsoft, Meta, OpenAI, and NVIDIA, and one E2E model trained on a publicly available Dutch speech corpus, i.e., CGN~\cite{oostdijk-2000-spoken}. 
The eight ASR models that we tested on DRES are the following. \textit{Google Chirp 3} is employed since it is the most advanced Google general purpose ASR model~\cite{google_speech_models} and the latest version of the Chirp model~\cite{zhang2023google} (January 2026). In addition, \textit{Google Telephony} was employed since it is optimized for telephone speech~\cite{google_speech_models}, making it suitable for spontaneous speech recognition. The third model is \textit{Microsoft Azure ASR model}~\cite{microsoft_speech_models}. We ran all three models using synchronous recognition through the API. All three are multilingual models and we used the Dutch language setting. 

Additionally, we downloaded five models from Hugging Face: the \textit{Massive Multilingual Speech} model~\cite{pratap2024scaling} developed by Meta was pre-trained with 491k hours of speech in 1406 languages and fine-tuned on 1162 languages~\cite{pratap2024scaling}, with the goal to recognize speech in diverse languages. We used greedy search decoding following the original implementation in~\cite{facebook_mms1ball}. 

The \textit{Whisper} model, developed by OpenAI, was initially trained on 680k hours of multilingual speech, mostly English~\cite{radford2023robust}. Whisper-large-V3 is the latest large Whisper model trained with even more training data and achieved SotA ASR performance across multiple languages (including Dutch) and accents~\cite{openai_whisper_large_v3}. Whisper-large-V3-Turbo achieves comparable performance with Whisper-large-V3 on Dutch speech from the FLEURS dataset~\cite{fleurs2022arxiv} but is optimized for faster inference~\cite{openai_whisper_discussion2363}. 
We set the Whisper parameter task to ``transcribe'', the language to ``Dutch'', and the temperature parameter to 0. 

\textit{NeMo-nl} is a Dutch ASR model developed by NVIDIA. We selected it since NeMo models were recently used to evaluate the effect of SE algorithms on speech intelligibility~\cite{Giraldo2025}. It was trained on 621 hours of speech from three publicly available datasets~\cite{nvidia_fastconformer_nl, nvidiacanary2024}: (1) Common Voice 12 (40 hours)~\cite{ardila2019common}, consisting of read speech; (2) Multilingual LibriSpeech (547 hours)~\cite{pratap2020mls}, consisting of read speech; and (3) VoxPopuli (34 hours)~\cite{wang2021voxpopuli}, consisting of parliamentary debates.

\textit{CGN-pre-trained Conformer}~\cite{zhang_dutchcgnconformerfbank} is a Dutch ASR model trained on approximately 700 hours of read and human-machine interaction (HMI) speech from CGN~\cite{oostdijk-2000-spoken}. It achieves SotA performance on the CGN original test sets and the full Jasmin-CGN set~\cite{cucchiarini2006jasmin, zhang2026speech}. For Whisper, NeMo-nl, and CGN-pre-trained-Conformer, we use beam search decoding with a beam size of 10.

\subsection{Evaluation metrics}\label{sec_eval}
ASR performance is measured in Word Error Rate (WER), ignoring non-linguistic symbols in the transcriptions. For ASR performance comparison, we conducted statistical significance test using a paired nonparametric bootstrap test~\cite{1326009, ferrer2024good} with 10,000 
speaker-based samples and 95\% confidence intervals (CIs). A difference is considered statistically significant if the CI excludes zero~\cite{ferrer2024good}. We
derived two-sided p-values from bootstrap samples by generating the null centered distribution, calculated as the proportion of samples where the absolute difference was at least as large as the observed absolute difference~\cite{BoosStefanski}. 

Speech quality is objectively evaluated using DNSMOS P.835 \cite{Reddy2022}, which is a no-reference objective speech quality measure used for estimating the MOS of enhanced speech and is commonly used in recent literature. The DNSMOS score ranges from 1 (very poor quality) to 5 (excellent quality). DNSMOS was trained to predict subjective MOS ratings obtained following ITU-T Rec. P.835 \cite{Reddy2022, ITUTP835}. 

\section{Results}
\subsection{Speech quality}
\label{subsec:SQ}

\begin{figure}[ht]
    \centering
    {\includegraphics[width=1.0\columnwidth, trim={0cm, 0.8cm, 0cm, 0.6cm}, clip]{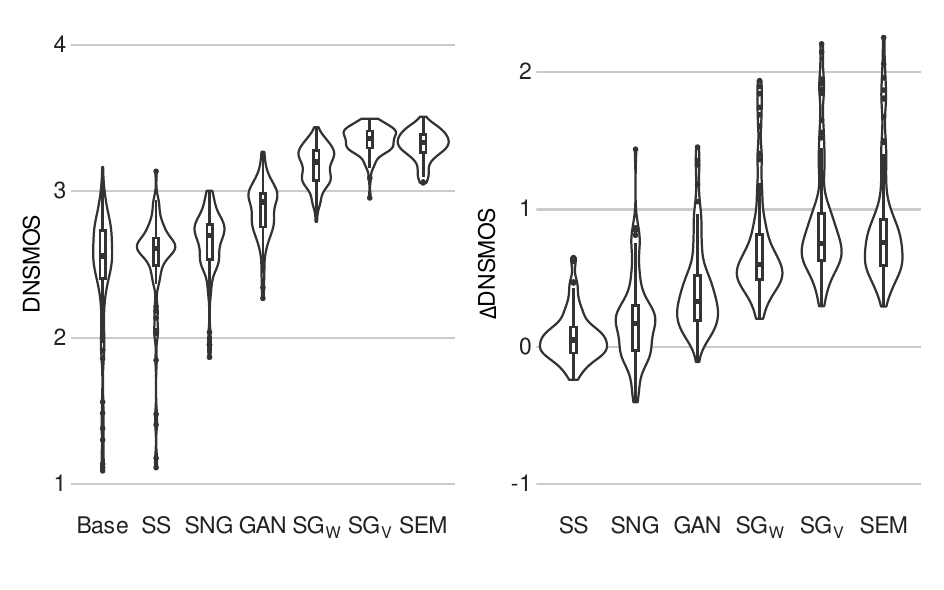}}
    \caption{\textbf{Left:} Estimated distribution of DNSMOS 
     for the baseline signal and  each of the SE algorithms. 
     Higher values imply a better overall quality.   \textbf{Right:} DNSMOS score improvements with respect to the baseline for each of the SE algorithms.}\label{fig_quality_enh}
\end{figure}

The left panel of Figure \ref{fig_quality_enh} shows the distribution of DNSMOS scores of the baseline signal (note that ``Base" is identical to ``Overall" in Figure \ref{fig_quality_loc}), and for each of the SE algorithms. As the figure shows, the median MOS-score of each SE algorithm improves compared to the baseline signal. The largest improvements are achieved for SG$_V$ and SEM. In the right panel of Figure \ref{fig_quality_loc}, the distribution of the improvement in DNSMOS with respect to the baseline signal is shown. Here it can be observed that SEM and both versions of SGMSE+ improve the speech quality for each utterance in the dataset. The other SE algorithms, in particular SNG and SS, reduce the speech quality for some of the utterances in the dataset. 

\subsection{ASR performance}
\begin{table}[ht]
    \centering
    \caption{Average WERs (\%) over the four locations of the
    eight ASR models on DRES before and after applying the 
    six SE algorithms. Bold denotes the lowest (best) WER in each column.}\label{tab: ASR-single}
    \resizebox{\columnwidth}{!}{
    \begin{tabular}{lccccccc} 
        \toprule
& \textbf{Base} & \textbf{SS}  & \textbf{SNG}  & \textbf{GAN} &\textbf{SG$_W$} & \textbf{SG$_V$} & \textbf{SEM} \\ 
        \midrule
        \textbf{Chirp} &\textbf{11.2} &\textbf{11.9}&\textbf{16.5}&\textbf{13.1}&\textbf{12.4}& \textbf{15.3} & \textbf{12.9} \\
        \textbf{Telephony} &18.3 &18.6&24.9&19.7& 19.1 & 21.3 &18.9\\
        \textbf{Azure} &20.8 &22.2&36.2&23.1 &20.7 &24.2 &23.5 \\
        \textbf{MMS} & 28.5 &31.0 &41.5& 32.5 &31.0 & 33.6 & 30.4\\
        \textbf{Whisper} &15.8 &16.3&20.1&17.8 &16.5 &18.7 & 16.0\\
        \textbf{Turbo} &62.5 &62.9&64.6&63.5&62.8 & 64.1 & 62.6\\
        \textbf{Nemo-nl} &33.4 &36.9&50.1&41.0 &36.1 &41.5 & 37.6\\
        \textbf{Conformer} &21.6 &27.1&94.7&27.7&24.0& 28.2 & 24.8\\
        \bottomrule
    \end{tabular}
    }
\end{table}

Table~\ref{tab: ASR-single} presents the average ASR results over the four locations on both the original and enhanced speech signals after applying the six SE algorithms, for each of the eight ASR models. For the original speech (see row ``Base''), the highest performance was obtained by Google Chirp 3 (Chirp). Chirp performed significantly better than the second-best model Whisper-large-V3 (Whisper), with an absolute WER difference of 4.60\% (95\% CI, [+3.95\%, +5.25\%], $p<.001$). 
Google Telephony (Telephony), Microsoft Azure ASR (Azure), and CGN-pre-trained Conformer (Conformer) performed relatively well on DRES speech, with average WERs around 20\%. Notably, Chirp was trained with far less data than Telephony and Azure. In contrast, Massive Multilingual Speech (MMS), Nemo-nl, and Whisper-large-V3-turbo (Turbo) showed poorer performance on DRES speech, especially Turbo, with the highest average WER of 62.5\%. 

As shown in Table~\ref{tab: ASR-single}, for four of the eight ASR models (Chirp, MMS, Nemo-nl, and Conformer), all six SE methods significantly degraded ASR performance compared to the baseline speech (Chirp: $p<.01$; MMS, Nemo-nl, and Conformer: $p<.001$).  
For Telephony and Whisper, SS had no significant effect. SEM had no significant effect on Whisper, but significantly degraded performance for Telephony ($p<.05$). The remaining SE methods significantly degraded ASR performance for both models (Telephony: $p<.001$; Whisper: $p<.05$). For Azure and Turbo, SG$_W$ did not improve nor degrade ASR performance. SEM had no significant effect on Turbo, but significantly degraded performance for Azure ($p<.001$). The remaining SE methods significantly degraded ASR performance (Azure and Turbo: 
$p<.05$).

\begin{figure}[ht]
    \centering
    {\includegraphics[width=0.97\columnwidth, trim={0 .31cm 0 .31cm}, clip]{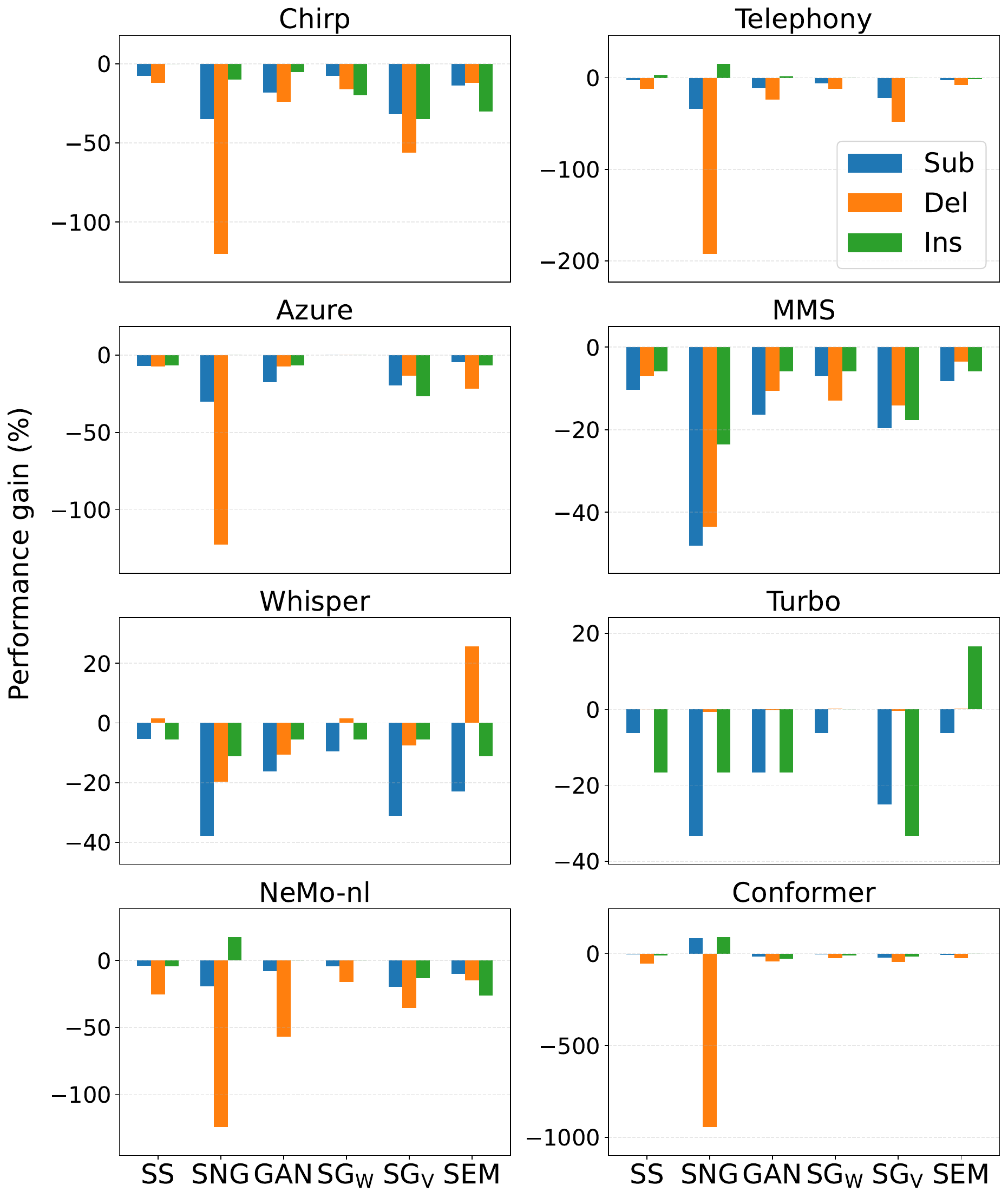}}
    \caption{Relative ASR performance gain after applying the six SE algorithms for the eight models on the baseline speech signal. Gains were calculated separately for substitution, deletion, and insertion error rates. Negative values indicate degradation.
    }\label{fig:performancegain}
\end{figure}

To further investigate the degradation of ASR performance after application of the different SE algorithms, Figure~\ref{fig:performancegain} shows the relative performance gains compared to the baseline signal in terms of substitution, deletion, and insertion errors for all eight ASR models. Negative numbers indicate a deterioration and positive numbers an improvement. For six of the ASR models (excluding Whisper and Turbo), five SE algorithms (excluding SEM) generally led to more substitution (blue bars) and deletion (orange bars) errors. We expect that this can be attributed to the SE algorithms suppressing low-energy speech components, leading to an increase in deletion errors, and to a distortion of phonetic cues, leading to an increase in substitution errors. In contrast, Turbo showed little or no degradation in deletion errors after SE (excluding SEM), but its insertion errors increased by $\sim$17\%–33\%. This observation can be related to the already high deletion error rate of Turbo on the baseline speech (57.1\%), which left less room for further degradation in deletion errors. In contrast, SEM reduced deletion errors for Whisper and insertion errors for Turbo, though substitution errors increased for both. These results indicate that the effects of SE are somewhat ASR model dependent. For most models, SNG and SG$_V$ led to substantial performance degradation, followed by SS, GAN, and SG$_W$. SNG increased deletion errors by more than 100\% for five out of the eight models (Chirp, Telephony, Azure, Nemo-nl, Conformer).  \looseness=-1

Table~\ref{tab:ASR-single-specific} splits the ASR results on both the original and enhanced speech signals for the four locations.
The results were highly similar across buildings: in line with the average ASR results (see Table~\ref{tab: ASR-single}), Chirp consistently achieved the best performance for both the original and enhanced speech, while Whisper generally showed the second-best performance among the eight ASR models, followed by Telephony, Azure, and Conformer. Overall, the performance differences among ASR models were broadly consistent across the four recording locations. 
For each ASR model and location, comparing the WERs before and after applying SE algorithms showed that none of the SE algorithms yielded statistically significant ASR performance improvements over the baseline speech.

\begin{table}[ht]
\centering
\caption{WERs (\%) of the 
eight ASR models on DRES recorded at the 
four locations before and after applying the 
five SE algorithms. 
Each block corresponds to one recording location. Bold denotes the lowest (best) WER in each column per recording 
location.}
\label{tab:ASR-single-specific}
\resizebox{\columnwidth}{!}{
\begin{tabular}{lccccccc}
\toprule
& \textbf{Base} & \textbf{SS} & \textbf{SNG} & \textbf{GAN} & \textbf{SG$_W$} & \textbf{SG$_V$} & \textbf{SEM} \\
\midrule
\multicolumn{8}{c}{Speech recorded at building \A}\\
\midrule
\textbf{Chirp}  & \textbf{9.8}  & \textbf{11.5} & \textbf{14.9} & \textbf{13.0} & \textbf{13.6} & \textbf{20.2} & \textbf{13.3}  \\
\textbf{Telephony}  & 17.0 & 17.4 & 28.0 & 19.1 & 19.7 & 26.5 &18.5 \\
\textbf{Azure}  & 17.1 & 21.7 & 28.3 & 20.7 & 17.1 & 27.3 &19.8\\
\textbf{MMS} & 29.9 & 34.7 & 46.0 & 36.3 & 35.5 & 42.7 & 33.3 \\
\textbf{Whisper}  & 14.0 & 15.8 & 19.1 & 17.2 & 18.4 & 25.6 &16.2\\
\textbf{Turbo} & 62.1 & 62.3 & 65.2 & 64.6 & 63.0 & 67.3 & 61.6\\
\textbf{Nemo-nl}  & 34.6 & 40.9 & 51.3 & 46.6 & 38.9 & 47.6 &40.8\\
\textbf{Conformer}  & 23.8 & 33.2 & 94.5 & 34.8 & 31.1 & 38.9 &31.7 \\
\midrule
\multicolumn{8}{c}{Speech recorded at building \B}\\
\midrule
\textbf{Chirp}  & \textbf{10.1} & \textbf{10.4} & \textbf{14.1} & \textbf{11.8} & \textbf{10.9} & \textbf{12.7} & \textbf{11.1}\\
\textbf{Telephony}  & 16.5 & 16.8 & 21.3 & 17.9 & 17.0 & 19.0 & 17.0\\
\textbf{Azure}  & 21.9 & 21.5 & 34.1 & 23.7 & 21.9 & 23.2 &23.0  \\
\textbf{MMS} & 25.3 & 25.9 & 34.8 & 28.2 & 26.1 & 28.2 & 25.8\\
\textbf{Whisper}  & 14.4 & 14.9 & 17.2 & 15.6 & 14.3 & 16.1 & 13.9  \\
\textbf{Turbo} & 66.0 & 66.3 & 67.2 & 66.6 & 66.0 & 66.4  &66.0\\
\textbf{Nemo-nl}  & 28.0 & 32.5 & 45.5 & 35.8 & 31.1 & 36.4 &33.3\\
\textbf{Conformer}  & 19.9 & 24.4 & 91.8 & 24.4 & 21.7 & 23.2 & 20.9\\
\midrule
\multicolumn{8}{c}{Speech recorded at building \C}\\
\midrule
\textbf{Chirp}  & \textbf{11.9} & \textbf{12.8} & \textbf{17.1} & \textbf{13.7} & \textbf{12.7} & \textbf{15.1} &  \textbf{13.3}\\
\textbf{Telephony}  & 19.0 & 19.4 & 25.8 & 20.3 & 20.0 & 20.8 &19.4 \\
\textbf{Azure}  & 23.8 & 26.1 & 39.6 & 24.9 & 23.7 & 25.5 &26.9\\
\textbf{MMS} & 30.6 & 32.9 & 42.0 & 33.6 & 32.6 & 33.3 &31.6\\
\textbf{Whisper}  & 16.6 & 17.1 & 20.6 & 18.5 & 17.0 & 18.5 &16.8\\
\textbf{Turbo} & 61.9 & 62.3 & 63.2 & 62.5 & 62.0 & 62.9&62.3 \\
\textbf{Nemo-nl}  & 35.3 & 37.1 & 49.3 & 41.6 & 37.1 & 42.0&38.6 \\
\textbf{Conformer}  & 22.4 & 27.6 & 94.4 & 27.9 & 23.9 & 27.7&24.2  \\
\midrule
\multicolumn{8}{c}{Speech recorded at building \D}\\
\midrule
\textbf{Chirp}  & \textbf{11.9} & \textbf{12.6} & \textbf{18.7} & \textbf{13.5} & \textbf{13.0} & \textbf{15.9}&   \textbf{13.9} \\
\textbf{Telephony}  & 19.6 & 19.9 & 26.0 & 21.0 & 19.8 & 22.1&20.1 \\
\textbf{Azure}  & 18.0 & 19.1 & 37.6 & 21.6 & 18.0 & 22.7&21.8 \\
\textbf{MMS} & 28.9 & 32.2 & 45.2 & 33.9 & 32.2 & 35.5&32.3  \\
\textbf{Whisper}  & 16.8 & 16.8 & 22.5 & 19.3 & 17.2 & 18.8&17.0 \\
\textbf{Turbo} & 60.1 & 60.7 & 63.6 & 61.3 & 60.6 & 62.2 &60.4 \\
\textbf{Nemo-nl}  & 35.9 & 39.2 & 54.4 & 43.1 & 38.4 & 43.4 &39.3 \\
\textbf{Conformer}  & 21.4 & 26.9 & 97.8 & 28.0 & 23.7 & 29.3& 26.3 \\
\bottomrule
\end{tabular}
}
\end{table}
Overall, our results indicate that these SE algorithms when applied to Dutch natural speech in noise degrade the performance of state-of-the-art ASR models; one reason could be due to the artifacts introduced by the SE algorithms. Furthermore our results show that a lower speech quality of the baseline signal does not necessarily correspond to worse ASR performance. While the speech recorded at \A had the lowest MOS score (see Fig.~\ref{fig_quality_loc}), a Kruskal-Wallis test indicated that the eight ASR models did not differ significantly in performance on speech recorded at \A-\D buildings (e.g., for Chirp, $\mathcal{X}^2(3)=2.60, p=.458$). A similar speech quality and ASR performance mismatch was observed for the enhanced speech. First, none of the SE algorithms yielded improved recognition performance. Second, as shown in Fig.~\ref{fig_quality_enh}, while the MOS score of the enhanced speech using SG$_V$ is significantly higher than that of SG$_W$ (paired Wilcoxon signed-rank test: $V=0$, $p<.001$, median difference $=0.155$), across all eight ASR models, the WER on SG$_W$-enhanced speech is significantly lower than SG$_V$-enhanced speech ($p<.001$).

\section{Discussion and conclusions}
In this work, we presented DRES: a dataset of Dutch elicited speech in realistic conditions. We evaluated eight state-of-the-art (SotA) ASR models on speech from DRES and evaluated the effect of several classical and modern single-channel speech enhancement (SE) algorithms. We found that, without SE, three out of eight ASR models, Google Chirp 3 (Chirp), Whisper-large-V3 (Whisper), and Google Telephony (Telephony) performed well on Dutch elicited speech in realistic background noise, achieving WERs of 11.2\%, 15.8\%, and 18.3\% respectively. Two models, Microsoft Azure (Azure) and CGN-pre-trained Conformer (Conformer), performed moderately on DRES, with WERs around 20\%. 
The remaining three ASR models (Multilingual Massive Speech (MMS), Whisper-larger-V3-turbo (Turbo), and Nemo-nl), performed significantly worse compared to the other five models ($p<.001$), with WERs substantially exceeding 20\%. 

To put these WERs into perspective: a recent study on relatively clean Dutch human-machine interaction (HMI) speech from diverse speaker groups (non-native accented, regionally accented speech and child speech) from  JASMIN-CGN reported WERs of approximately 23\% for Google Chirp 2 and Google Telephony~\cite{zhang2026speech}, which are higher than the WERs of the best-performing ASR (Google Chirp 3) on DRES. Similarly, Azure, MMS, Whisper-large-v3, NeMo-nl, and the CGN-pre-trained Conformer also had higher WERs on JASMIN-CGN HMI speech than on DRES, suggesting that these systems are more robust against acoustic variability due to realistic background noise than due to demographic differences between speakers. Note that Whisper-large-v3-turbo achieved a WER of approximately 32\% on the clean HMI speech, which was substantially lower than its WER on DRES~\cite{zhang2026speech}. 

Interestingly, none of the SE algorithms improved ASR performance on DRES, irrespective of the ASR model. Instead, the SE algorithms generally increase the deletion and substitution errors for most ASR models. This is in contrast to e.g.\cite{Lemercier2023, Yang2026}, who found that SGMSE+ improved E2E ASR performance on noisy English speech. In our experiments on DRES, SGMSE+ degraded performance for six out of eight ASR models and did not improve any. We hypothesize that the lack of improvement in recognition performance may be attributed to language differences and to different noise scenarios: where \cite{Lemercier2023, Yang2026} used artificial English noise-speech mixtures for read speech, we used noisy Dutch speech recorded in real-world settings. Our findings, however, extend prior work, where single-channel SE failed to improve HMM-GMM and hybrid DNN-HMM ASR performance~\cite{fujimoto2019, Iwamoto2022, 6296526, Wang2020, Ho2023}, to SotA E2E ASR models. These results demonstrate the importance of assessing ASR and SE performance across diverse and realistic scenarios. 

In summary, DRES has clear potential to benefit both the ASR and SE communities. Our corpus of realistic elicited Dutch speech in diverse natural background noise enables evaluation of SotA ASR models on realistic speech and allows for assessment of SE algorithms on natural speech in noise, thus moving beyond artificial noise mixtures. Our results showed that our tested classical and SotA single-channel SE methods do not improve ASR performance on Dutch speech in realistic scenarios, suggesting that careful consideration is required when integrating SE with modern ASR systems, and motivates further research on SE for natural speech in noise. In future work, we plan to extend our work by including statistical SE algorithms, left out in this work, and to investigate the effect of multi-channel SE approaches, which exploit spatio-temporal information to enhance speech. 
\section{Acknowledgements}
\ifcameraready
The authors thank Ilse Huisman and Ansen Weng for their support with constructing the ground-truth transcriptions.

Generative AI (ChatGPT) was used to improve language and readability, and debug Python scripts for bootstrapping.
\fi

\bibliographystyle{IEEEtran}
\bibliography{IEEEabrv,mybib}

\end{document}